\documentstyle[aps,prl,multicol]{revtex}
\begin{document}
\draft
\preprint{\today}
\title{Impurity correlations in dilute Kondo alloys}
\author{Victor Barzykin}
\address{National High Magnetic Field Laboratory,
1800 E. Paul Dirac Dr., Tallahassee, FL 32310}
\author{Ian Affleck\footnote{Permanent address: Canadian Institute for
Advanced Research and Department of Physics 
and Astronomy, University of British Columbia, Vancouver, BC,
Canada, V6T 1Z1}}
\address{Institute for Theoretical Physics,
University of California Santa Barbara, CA 93106-4030}
\maketitle
\begin{abstract}
The single impurity Kondo model is often used to describe metals
with dilute concentrations ($n_i$) of magnetic impurities.  Here we
examine how dilute the impurities must be for this to be valid 
by developing a virial expansion in impurity density.  The  
$O(n_i^2)$ term is determined from results on the 2-impurity
Kondo problem by averaging over the RKKY coupling.  The 
non-trivial fixed point of the 2-impurity problem could
produce novel singularities in the heat capacity of
dilute alloys at $O(n_i^2)$.
\end{abstract}
\begin{multicols}{2} 
\narrowtext

\section{Introduction}

The physics of the single-impurity Kondo model, the problem describing
a magnetic moment interacting with a sea of conduction electrons, 
is well understood\cite{hewson}. At high temperatures the moment is free.  
The spin susceptibility of the impurity obeys a simple Curie-Weiss law, and
there is a large additional term in the entropy, $S=k_B \ln 2$. 
As the temperature decreases, the impurity spin disappears as it 
forms a singlet with conduction electrons. The characteristic 
energy scale at which this happens is the Kondo temperature,
which is exponentially small at small couplings,
$T_K \sim E_F exp(-1/(J \rho))$, where $J$ is the $s-d$ exchange 
constant, $\rho$ is the density of states. 
The $1/T$ divergence of the spin susceptibility is then effectively
cut off by this scale at low temperatures.
The singlet state formed by conduction electrons and impurity
can be described at low temperatures $T \ll T_K$ as a Fermi liquid 
with enhanced density of states, $\delta \nu \sim n_i/T_K$, where
$n_i$ is the concentration of impurities.

The single-impurity Kondo model is often used as a way to
obtain insights into the nature of the ground state for the 
Kondo lattice problem,
a toy model for a number of rare earth compounds, such as heavy fermions.
Enhancement of the density of states due to the Kondo physics explains 
qualitatively the heavy masses in the heavy fermion materials. 
The local moments in a Kondo lattice
interact through the electron-mediated RKKY interaction, which
competes with the Kondo physics. The low temperature behavior
of the Kondo lattice is often thought to be determined by this competition.
Conventional mean field models
of the heavy fermions\cite{coleman}, however, often neglect 
the RKKY interaction, since it only appears as a correction to the mean field
behavior, and therefore is difficult to calculate. 

The simplest model that captures the physics of this competition 
is the two-impurity Kondo 
model\cite{jkw,varma,hf}. Depending on the relation between the RKKY
interaction and the Kondo temperature and the presence of particle-hole
symmetry breaking, the impurity spins can each get get compensated 
by conduction electrons, form a singlet with each other, or
a combination of both.  A surface of Fermi liquid fixed points occurs
in the problem as well as an unstable non-trivial critical point 
associated with critical behavior of the  
thermodynamic functions \cite{jv,jkm,mkj,al,gan,svk}. 

Despite extensive theoretical work on the 2-impurity problem, we are
not aware of any effort to apply the results in a systematic way
to experiments.  The appropriate way to do this, for dilute alloys, 
was pointed out in 1970 by Larkin and collaborators \cite{lh}.
One should develop an expansion of thermodynamic quantities in
powers of the impurity concentration, $n_i$, a virial expansion.
The pioneering work on this subject \cite{lh} only considered
the high-$T$ limit where the Kondo interaction could be ignored
and only the RKKY interaction considered.  At $T$ of $O(T_K)$
or lower the Kondo interaction must be included.  The
$O(n_i^2)$ term in the virial expansion is completely determined
by properties of the 2-impurity problem appropriately averaged
over the strength of the RKKY interaction.  This calculation
gives an estimate of the characteristic impurity concentration at
which the single impurity Kondo model breaks down.  
 Analysis of the 
$N$-impurity models with $N>2$ could provide a further insight into the structure of
this expansion. 

We note that 
deviations from linear dependence of thermodynamic quantities on
impurity concentration do not arise solely from the mechanism considered here.
The Kondo temperature depends exponentially on the exchange integral, and changes
in the composition of an alloy can alter it significantly.
Such corrections, however, can be incorporated in the single-impurity picture,
even though they alter the linear dependence of the thermodynamic and magnetic properties
of a dilute alloy on concentration.

In the next section we discuss the subtle issue of length scales in
the Kondo problem.  Section III sets up the virial expansion,
following \cite{lh}.  Section IV reviews some relevant results
on the 2-impurity Kondo problem.  The $O(n_i^2$) term is then
considered for $T \gg T_K$ in Section V and at $T\leq T_K$ in
Section VI.  Quantitative results would certainly require accurate
numerical solution of the 2-impurity problem for all values or
the RKKY coupling.  Here we just apply the large-N approximation
and make some qualitative remarks on the effect of the
non-trivial critical point.  Section VII contains conclusions.

\section{Length scales in the Kondo problem}
There are (at least) 3 different ways of estimating 
a lower bound on the average separation
of impurities neccessary for the single impurity model to be valid.  
One could require that each impurity be far from the others compared
to the size $\xi_K$ of its screening cloud.  Alternatively,one could require that
the density of electron states within an energy $T_K$ of the Fermi
surface be at least as large as $n_i$ (the Nozi\`eres exhaustion principle).
This appears to be a neccessary (although not obviously sufficient) condition 
for the screening cloud wave-functions
 from each impurity to form an approximately orthogonal set.  Or, finally, one could
require that the average RKKY interaction be small compared to $T_K$.

As a mechanism of Kondo screening, the picture of
spin exhaustion cloud of exponentially large size 
$\xi_K \sim v_F/T_K$ ($v_F$ is the Fermi velocity) has often been 
adapted\cite{hewson}. 
From the RG point of view, clearly, this length scale is present
in the problem, since the low-energy Fermi liquid theory\cite{Nozieres} is only valid
for $k-k_F \ll T_K/v_F$, or at distances $r \gg \xi_K$ away from the 
impurity. Direct calculations of the screening cloud 
profile\cite{BA} for the spin-spin correlators of the single-impurity
Kondo model demonstrate that the screening cloud of conduction
electrons indeed forms at the distance scale $\sim \xi_K$. Note that 
this is a more dynamical type of a screening than that which occurs for
charge impurities in a Fermi liquid since it involves a linear combination of 
states where the impurity spin and the screening electron spin are in either
an up-down or down-up configuration. In particular, the finiteness of the
susceptibility at $T \rightarrow 0$ should not be attributed to a static 
conduction electron polarization cancelling the impurity  spin polarization.
Rather it results from the tendency of the impurity to form a singlet with 
the screening electron. This tendency is very well illustrated by a
calculation of an equal-time correlator, which provides an instant snapshot of the 
system, and the zero-frequency spin correlator, or the spin susceptibility\cite{BA}.
Indeed, it has been shown that the magnetic moment of conduction electrons exactly 
compensates that of the impurity at the same moment in time. Yet, there is
no net polarization of the conduction electrons (as illustrated by a calculation
of zero-frequency spin susceptibility), the fact also known as the 
Anderson-Clogston theorem.

Existence of an exponentially large length scale could have potentially
strong effect on the theory of alloys with magnetic impurities. Indeed, 
typical $T_K \sim 10K$ and $E_F \sim 10 eV$ makes $\xi_K \sim 10000a$,
where $a$ is the lattice spacing, much larger than typical distance 
between two impurities. This issue was addressed in one dimension for
Luttinger liquids with magnetic impurities\cite{ZKE}, where it was found
that the crossover happens at $n_i \sim 1/\xi_K$. At higher dimension,
however, no such crossover was found experimentally. The physical reason
for this is that the screening cloud wave function has an oscillatory character. 
Let us make a simple estimate of the overlap of the screening clouds 
from two separate impurities, with one impurity located at a distance $R$ 
apart from the other. 
We shall make no particular assumptions about the
screening cloud wave function except that it is made up of Fourier modes with 
$|k-k_F|$ less than or of order of $1/\xi_K$. Thus the overlap is:
\end{multicols}
\widetext
\begin{equation} O(R) \equiv \int d^3 r\psi^* (|\vec r+\vec R/2|)\psi (\vec r-\vec
R/2|) =\int {d^3 k\over (2\pi)^3}e^{i\vec k\cdot \vec R}|\psi
(k)|^ 2=\int_0^\infty dk k^2|\psi (k)|^2{\sin (kR)\over 2\pi^2kR}
\end{equation}
\begin{multicols}{2}
\narrowtext
So far we have only used the spherical symmetry of the screening cloud
wave-function and hence its Fourier transform.  Now we assume:
\begin{equation}
|\psi (k)|^2 \approx (\xi_K/k_F^2)f[(k-k_F)\xi_K],
\end{equation}
where the scaling function $f(y)$ obeys the normalization condition:
\begin{equation}
\int dy f(y)=2\pi^2,
\end{equation} in order that $O(0)=1$.  Thus we get:
\begin{equation}O(R) = (1/2\pi^2)\int dy f(y)\sin [k_FR+(R/\xi_K)y]/k_FR.
\end{equation}
For $R \ll \xi_K$ this reduces to:
\begin{equation} O(R) = \sin (k_FR)/k_FR. \end{equation}
Thus the overlap is small for $k_FR \gg 1$.
This calculation can be easily extended to the 1 or 2 D case.  For 1 D the
overlap is essentially $\cos (2k_FR)$ for $R \ll \xi_K$.  It doesn't get
small until $R \gg \xi_K$.  In 2 D it is a Bessel function, $J_0(k_FR)\approx
\sqrt{2/\pi k_FR}\cos (k_FR-\pi /4)$ (for $k_FR \gg 1$), again becoming small
for $k_FR \gg 1$. We see that in dimensions higher than one the overlap of the
two screening cloud wave functions is suppressed by 
$1/\sqrt{k_F R}$(2D) or $1/k_F R$(3D), 
small factors at large enough inter-impurity separation. In $1D$ this overlap is $O(1)$.
Thus the dimensionality of the problem could be the main reason why the large Kondo scale
is absent in alloys.

The well-known
Nozi\`eres' exhaustion principle states that for each impurity spin there should be
a conduction electron in the vicinity $\sim T_K$ of the Fermi energy that
screens it. This produces the following estimate for the inter-impurity separation:
$R \sim (\xi_K k_F^{-2})^{1/3}$ in 3D, $R \sim \sqrt{\xi_K k_F^{-1}}$ in 2D,
and $R \sim \xi_K$ in 1D. 

On the other hand, one can argue that the single-impurity
model stops working when the RKKY interaction at the average inter-impurity distance
becomes comparable to the Kondo temperature, $J_{RKKY} \sim T_K$. This gives a somewhat
higher estimate for the concentration at which one needs to account multi-impurity effects
(or lower estimate for the average inter-impurity distance),
$R \sim (\lambda^2 \xi_K k_F^{-2})^{1/3}$ in 3D,
$R \sim \sqrt{\lambda^2 \xi_K k_F^{-1}}$ in 2D,
and $R \sim \lambda^2 \xi_K$ in 1D, because of its dependence on the small 
coupling constant,
$\lambda \equiv (J \rho)$.

\section{The virial expansion.}

We start with the ordinary s-d Hamiltonian for a collection of magnetic impurities in
an electron gas located at arbitrary positions in space:
\begin{equation}
H_{sd} = - {J \over 2N}   \sum_{j,{\bf k},{\bf k'}} a^{\dagger}_{{\bf k'} \alpha}
(\bbox{\sigma}_{\beta}^{\alpha} \cdot {\bf S}^j_{imp}) a_{{\bf k}}^{\beta} 
e^{i({\bf k} - {\bf k'}){\bf r}_j}
\label{sd}
\end{equation}
The free energy for $N$ impurities is determined by
the formula:
\begin{equation}
F_{1,2,\cdots,N} = -T \ln{\langle \exp\{-\int_0^{1/T} H_{sd} (\tau) d \tau\} \rangle}
\label{Freen} 
\end{equation}
To obtain a virial expansion, one can introduce the quantities $f$ defined by recurrence
relation:
\begin{eqnarray}
f_i &=& F_i \nonumber \\
F_{ij \cdots n} &=& \sum_k f_k + \sum_{kr} f_{kr} + 
\cdots \sum_{kr \cdots m} f_{kr \cdots m} + f_{ij \cdots n}, 
\label{vir}
\end{eqnarray}
where the summation is carried out over different sets of indices ${ij \cdots n}$.
The function $f$ vanishes if the distance between any two impurities tends to infinity.
For example,
\begin{equation}
f_{ij} = F_{ij} - (F_i + F_j).
\end{equation}
Averaging Eq.(\ref{vir}) over the distribution of the impurities and going to
the thermodynamic limit $N \rightarrow \infty$, we obtain an expansion of the free energy
in powers of the density. For magnetic impurities interacting with an RKKY interaction such
expansion was carried out in Ref\cite{lh}. The $n_i^2$ term in the Kondo regime without the RKKY
term was calculated in Ref.\cite{gk}, although the influence of the RKKY interaction on 
this result was discussed in a later paper. The ordinary Kondo term is proportional to the
density of impurities, $n_i \equiv N_i/V$. The $n_i^2$ contribution takes the following form:
\begin{equation}
F^{(2)} = N_i {n_i \over 2} \int d {\bf R} (F_{2i}({\bf R}) - 2 F_{Kondo}).
\label{2coef}
\end{equation}
Here $F_{2i}({\bf R})$ is the free energy of the two impurities separated by a distance 
${\bf R}$, $F_{Kondo}$ is the usual single-impurity Kondo term. 
Thus the $n_i^2$ correction is competely determined by the two-impurity physics.
Eq.(\ref{2coef}) and
similar expressions for thermodynamic functions (derivatives of F) could
be useful for numerical determination of the temperature dependence of the $n_i^2$ term 
in thermodynamic functions from the results on the two-impurity model.

\section{The two-impurity Kondo model.} 

Before calculating the $n_i^2$ term let us discuss the relevant physics of the
two-impurity model. The study of the critical point for two impurities
interacting with the Hamiltonian Eq.(\ref{sd}) starts with a reduction of
a three-dimensional problem to a one-dimensional one\cite{jv,al,gan}. This is
done by introducing orthogonal $1D$ fermionic operators,
\end{multicols}
\widetext
\begin{equation}
\psi_{1,2}(k)=\int d \Omega {k \over \sqrt{2}}\left[{1 \over N_e(k)} 
cos({\bf k \cdot R}/2)
\pm {i \over N_o(k)} sin({\bf k \cdot R}/2)\right] a({\bf k}),
\end{equation}
\begin{multicols}{2}
\narrowtext
where we have integrated over spherical angle, and
\begin{equation}
N_{e,o}(k) = \sqrt{1 \pm {sin(kR) \over kR}}.
\end{equation}
Then the two-impurity model can be rewritten in $1D$ form.
While the transformation itself is exact, the $k$-dependent couplings arise.
Then these couplings are assumed to be $k$-independent for excitations
in the vicinity of the Fermi surface, an assumption questioned in Ref.\cite{fye}. 
The main drawback of such assumption is that the RKKY exchange interaction 
then does not have a correct oscillating
character.  The reason is that the main contribution to 
RKKY interaction comes from energies $\sim v_F/R$, where $v_F$ is the Fermi velocity,
$R$ is the distance between impurities. Apart from this, however, this procedure
is fully justified. 
One can view the process of
cutoff renormalization in two stages. During the first stage, the cutoff 
$\Lambda$ is reduced below $v_F/R$, $\Lambda \ll v_F/R$, 
and oscillating RKKY interaction appears as a result:
\end{multicols}
\widetext
\begin{equation}
H_{int} = {1 \over 2} \lambda 
\bar{\psi}_1^{\dagger}(0) \bbox{\sigma} \cdot {\bf S}_1 \bar{\psi}_1(0)+
{1 \over 2} \lambda \bar{\psi}_2^{\dagger}(0) 
\bbox{\sigma} \cdot {\bf S}_2 \bar{\psi}_2(0)
+ I {\bf S}_1 \cdot {\bf S}_2. 
\label{1D}
\end{equation} 
\begin{multicols}{2}
\narrowtext
\noindent
Here $\lambda = J \rho$ is the Kondo coupling constant and 
$I = V_0 cos(2 k_F R)/R^3$ is the usual RKKY term,
$V_0 = \rho J^2/(16 \pi)$. Note that both impurities are now at 
the same place in the $1D$ model.
During the second stage the $1D$ Hamiltonian Eq.(\ref{1D})
is renormalized. The wave functions $\bar{\psi}_1$ and
$\bar{\psi_2}$ are not orthogonal; they can be written in terms of orthogonal
$\psi_1$ and $\psi_2$ as follows:
\begin{eqnarray}
\bar{\psi}_1 &=& \alpha \psi_1  + \beta \psi_2 \\
\bar{\psi}_2 &=&  \beta \psi_1 + \alpha \psi_2,
\end{eqnarray}
where 
\begin{eqnarray}
\alpha &=& {N_e(k_F) + N_o(k_F) \over 2} \simeq 1, \\
\beta &=& {N_e(k_F)-N_o(k_F) \over 2} \simeq {sin(k_F R) \over 2 k_F R}
\end{eqnarray}
This orthogonal basis differs from the usual ``even'' and ``odd'' parity 
wavefunctions. It is seen explicitly in this basis that for
$k_F R \gg 1$ channel mixing terms play minor role and can be neglected.
Other marginal operators appear in the interaction Hamiltonian Eq.(\ref{1D}).
Those, however, are suppressed by factors of $k_F R$, and therefore can 
be neglected.  

\section{High-temperature results.}

We can now analyse the $n_i^2$ corrections in a Kondo alloy of spin-$1/2$
impurities using the $1D$ Hamiltonian 
Eq.(\ref{1D}). At high temperatures ($T \gg T_K$) this can be done by including the
RKKY term in Eq.(\ref{1D}) in the non-interacting Hamiltonian, and
perturbing in the Kondo interaction. Instead of averaging over distances as
in Eq.(\ref{2coef}) one can do averaging over the strength of the RKKY coupling
\cite{lh}. Indeed, the interaction $I(R)$ is a product of the smooth function
$R^{-3}$ times the rapidly oscillating function $cos(2 k_F R)$. It
is convenient to carry out the integration in formula Eq.(\ref{2coef}) by first
averaging the integrand over the period of oscillations. Changing the order of
averaging and integration, we obtain:
\begin{equation}
F^{(2)} = {4 N n_i V_0 \over 3 T} 
v. p. \int {dy \over y^2}  (F_{2i}(y,\lambda) - F_{2i}(0,\lambda)),
\label{Fr}
\end{equation}
where $y = I(R)/T$, $v.p.$ stands for ``principle value''. 
The behavior of the bulk spin susceptibility and the heat capacity of 
a Kondo alloy at temperatures $T \gg T_K$ is well known:
\begin{eqnarray}
\chi(T) &=& {(g \mu_B)^2 n_i \over 4 T}(1-\lambda_{eff}(T)) \nonumber \\
C(T) &=& {3 \over 4} \pi^2 n_i \lambda_{eff}(T)^4, 
\end{eqnarray}
where $\lambda_{eff}(T) \simeq \lambda + \lambda^2 \ln\left[{\Lambda \over T}\right]$
is the effective coupling constant at energy scale $T$.
To the lowest orders in the Kondo coupling we get the following result for the $n_i^2$ correction
to the heat capacity:
\begin{equation}
\delta C = {4 \over 3} {n_i^2 V_0 \over T} 
\left(1 - \lambda^2 \ln\left[{\Lambda \over T}\right]\right),
\label{Chigh}
\end{equation}
while for the magnetic susceptibility we find:
\begin{equation}
\delta \chi = \kappa (g \mu_B)^2 {n_i^2 V_0 \over 3 T^2}\left(1 - \lambda - 
2 \lambda^2 \ln\left[{\Lambda \over T}\right]\right),
\label{Chihigh}
\end{equation}
where
\begin{equation}
\kappa = v. p. \int_{- \infty}^{\infty} {d y \over y^2} {1 - e^y \over 3 + e^y} 
\simeq -0.501447,
\end{equation}
and $\Lambda \ll v_F/\bar{R} \simeq v_F (T_K/V_0)^{1/3}$ is the cutoff of 
the $1D$ model. Note that the small-bandwidth
effective theory of the two-impurity Kondo model is, of course,
only valid for calculating low energy properties. Thus the
above results are valid in the range $T_K \ll T \ll \Lambda$.
Since the RKKY interaction is relevant
at weak coupling, it gives rise to power law singularities in
thermodynamic functions at high temperatures\cite{lh}. 
The marginal Kondo interaction gives rise to weaker logarithmic
terms in Eqs.(\ref{Chigh}),(\ref{Chihigh}).
As we will see below in the next section, these singularities are cut off at
temperatures of the order of $T_K$.
The parameter of the virial expansion is $n_i V_0/T \ll 1$.
The major contribution to the $n_i^2$ term comes from distances 
$\bar{R} \sim (V_0/T)^{1/3}$,
where the amplitude of the RKKY term is of the same order as temperature.
The virial expansion fails at high enough densities when the average distance
between impurities becomes comparable to $\bar{R}$. Note that 
the second virial coefficients for the spin susceptibility and
the heat capacity Eqs.(\ref{Chigh}),(\ref{Chihigh}) cannot 
be written simply in terms of an effective coupling.
Averaging over distances in the virial expansion correspond to
different values of RKKY exchange $I$, which can be quite large.
It is clear that $I$ ruins the simple dependence of
physical quantities on $\Lambda$ and bare coupling.
Consider what happens when we take $I$ either very large or
very small, continuing to treat $I$ exactly and doing
perturbation theory in the Kondo couplings.  If $I$ is very
small (compared to $T$) we should be able to ignore it and
then we just get two decoupled single impurity Kondo problems.
In this limit we certainly recover the standard reduction of
$\Lambda$-dependence to renormalized coupling.  On the other
hand, if $T \ll |I|$ ($I<0$) then, at low energies, we have
essentially a two-channel $S=1$ Kondo problem.  Again we expect
a standard reduction of the dependence on $\Lambda$
corresponding to the beta function for that problem.
However, the evolution of effective Kondo coupling with
energy scale is different in these two limits (two
decoupled  one-channel $S=1/2$ Kondo problems versus one two-channel
S=1 Kondo problem).  Thus, as we lower our cut-off past $I$
the nature of the dependence on the cut-off changes.  This
is to be expected on general grounds.  When we
have a physical energy scale, like $I$ in the problem we don't
in general get simple dependence on the cut-off scale.  
In a massive quantum field theory effective coupling constants
are only useful at energy scales large compared to the mass.

\section{Corrections at low temperatures.}

Let us now consider this problem at low temperatures. Averaging
over the distances between two impurities involves integration over all RKKY couplings
(see Eq.(\ref{Fr})), which makes analytical analysis extremely difficult.
Instead let us consider first the $SU(N)$ generalization of the two-impurity model, 
for which explicit results for the linear term in the specific heat ($\gamma$) and the 
bulk magnetic susceptibility were obtained\cite{jkm} as a function of RKKY coupling strength.
For a $U = \infty$ $SU(N)$ generalized two-impurity Anderson model with RKKY interaction 
they find, in the large-$N$ limit:
\begin{equation}
\gamma=\pi^2 \chi = {N \pi \over 3} {2 e^{-B \delta} cos \delta \over (1 - A^2) T_K^*},
\label{1overN}
\end{equation}
where $A \simeq sin(k_F R)/k_F R$, $B \simeq cos (k_F R)/k_F R$, $\delta$ is related to
the scattering phase shift in even and odd channels, $\delta_{e,o} = \pi/2 \pm \delta$,
$N$ is the $SU(N)$ expansion parameter, $T_K^*$ is the Kondo temperature. $\delta$ is
found from the solution of appropriate mean field equations. To find the $n_i^2$ correction
to $\gamma$ and $\chi$, we have to average Eq.(\ref{1overN}) over distances, or over RKKY 
interaction. Since for the average distance $k_F \bar{R} \gg 1$, $B \rightarrow 0$, 
$A \rightarrow 0$. For these values of parameters the model has a first-order phase
transition at $(I/T_K)_{c}= 8/\pi$, with $\delta \simeq 0$ when $(I/T_K) < 8/\pi$ and
$\delta \simeq \pi/2$ if $(I/T_K) > 8/\pi$. We can apply the same averaging technique as in
Eq.(\ref{Fr}) (with $y = I(R)/T_K$):
\begin{equation}
\chi^{(2)} = {4 N n_i V_0 \over 3 T_K} 
v. p. \int {dy \over y^2}  (\chi_{2i}(y,\lambda) - \chi_{2i}(0,\lambda)).
\label{Fr1}
\end{equation}
Performing the integration, we easily find:
\begin{equation}
\gamma = \pi^2 \chi = N_i {\pi N \over 3 T_K} \left(1 - { \pi n_i V_0 \over 3 T_K} \right),
\label{gam}
\end{equation}
where the first term in Eq.(\ref{gam}) corresponds to the ordinary Kondo effect
in the $SU(N)$ model. One important result of the virial expansion is to find the
density at which the single-impurity approximation no longer works. 
As one can see from Eq.(\ref{gam}), this happens at $n_i \sim T_K/V_0$, or when
the Kondo temperature becomes comparable to the amplitude of the RKKY interaction
at the average inter-impurity distance. 

The mean field $N=\infty$ solution suffers from a number of defficiencies. In particular,
it was shown\cite{jv,svk,gan} that for two spin-$1/2$ impurities the quantum phase 
transition is second, not first order. The critical point for two spin-$1/2$ impurities 
is only present if the model is particle-hole-symmetric.
Under these conditions it was found\cite{svk,gan} that the specific heat coefficient
$\gamma$ diverges at 
the critical point as $|I/T_K - (I/T_K)_c|^{-2}$, $(I/T_K)_c \simeq 2.2$, while 
the uniform spin susceptibility $\chi$ does not. Let us find out whether this
divergence produces any singularity in the thermodynamic quantities. Unlike in
the $N=\infty$ model, where there is always a first order phase transition,
One can identify possible singularities at $y \simeq 0$ and 
$y \simeq y_c$ from Eq.(\ref{Fr1}).
At $y \equiv I/T_K \rightarrow 0$ the two Kondo impurities are independently screened.
Since impurity spin at the Kondo fixed point should be replaced by the local spin density,
\begin{equation}
{\bf S}_{1} \propto {v_F \psi_L^{\dagger}(0) \bbox{\sigma} \psi_L(0) \over T_K},
\end{equation}
so that the RKKY interaction can be written as follows:
\begin{equation}
H_{RKKY} \propto {I v_F^2 \over T_K^2}  (\psi_1^{\dagger}(0) \bbox{\sigma} \psi_1(0)
\cdot \psi_2^{\dagger}(0) \bbox{\sigma} \psi_2(0)).
\end{equation}
Perturbing in $H_{RKKY}$, one easily finds that the singularity in $\gamma$ and $\chi$
at $y \rightarrow 0$ is logarithmic, i.e. principle value-integrable. For $y \simeq y_c$
the bulk susceptibility $\chi$ does not have a singularity, whereas $\gamma$ has a power
law divergence $\sim (y-y_c)^{-2}$. The form of this singularity in the vicinity of $y_c$ 
was investigated ealier using bosonisation by Gan\cite{gan}, who showed that it is 
effectively cut off at finite temperatures at $\delta y \propto \sqrt{T}$. 
This produces a square root singularity in $\gamma$,
\begin{equation}
\delta \gamma \propto {N_i \over T_K} {n_i V_0 \over \sqrt{T T_K}}
\label{sing}
\end{equation}
Since $\chi$ does not diverge at the critical point, such singularity is absent for the
bulk susceptibility. It is known\cite{al} that in the absence of particle-hole symmetry there
are relevant perturbations near the zero-temperature critical point of the two-impurity
Kondo model. Such perturbations tend to wipe out the singularity at low temperatures,
so it is not clear whether this effect could be observed in real materials.

\section{Conclusions.}

To summarize, we have considered cluster expansion in ``dense'' Kondo alloys.
This expansion ties the two-impurity model, which has been investigated in detail,
to experiment.  At high temperatures $n_i^2$ corrections 
produce additional $1/T$ and logarithmic singularities in thermodynamic quantities.
At low temperatures the $n_i^2$ term becomes large when the RKKY interaction at
the average distance between impurities becomes of the same order as the Kondo temperature.
Since the parameter of expansion in density is $n_i/n_c$, where 
$n_c = T_K/V_0 \propto \lambda^{-2} \exp(-1/\lambda)$ 
($\lambda = \rho J$ is the dimensionless
Kondo coupling constant) is exponentially small, deviation from  the single-impurity behavior
could be observed in dilute alloys  with magnetic impurities. We have also seen that
the presence of the intermediate fixed point in the two-impurity problem could lead to a mild
low-temperature singularity in the specific heat, but not the bulk magnetic susceptibility.
Further numerical investigation of this problem is desirable. 
In some rare earth alloys, such as Yb$_x$Y$_{1-x}$CuAl\cite{hewson},
the behavior of thermodynamic functions is linear in $x$ up to
very large $x \simeq 0.9$. This observation could be related to the fact that these ions are
often described by the Coqblin-Schrieffer model (N=8 in case of $Yb^{3+}$), where the 
RKKY interaction is suppressed ($O(1/N^2)$\cite{Doniach}).

Our calculation of the second virial coefficient shows that 
the density of impurities at which multi-impurity effects become important
is {\em higher} than that given by Nozi\`eres' exhaustion principle. According
to this principle, at these densities ($n \lesssim n_c$) there are not enough states 
to screen all impurities. Since Nozi\`eres' principle is multi-impurity in nature, 
it may be necessary to study the convergence properties of the
expansion itself rather than the behavior of a given term to claim its failure. 

Finally, we note that the reason why the large Kondo length scale is absent 
in the problem is purely geometric. The channel mixing terms between 
two impurities are large in the 
one-dimensional problem. However, in $3D$ they become suppressed by a 
small factor of $1/k_F R$, and the physics of the two-impurity model (and related $n_i^2$ term
in the cluster expansion) is fully determined by the competition of the RKKY and Kondo 
effects.

We would like to thank D. J. Scalapino for discussions
and correspondence regarding the physics of the Kondo screening cloud. 
One of us( VB) is also greatful to V. Dobrosavljevic and J. R. Schrieffer for discussions
and to L. P. Gor'kov for critical comments and encouragement.
This work was supported in part (VB) by the National High Magnetic
Field Laboratory through NSF cooperative agreement 
No. DMR-9527035 and the State of Florida and in part (IA) by NSERC of Canada.

\end{multicols}
\end{document}